\documentclass[manuscript,screen]{acmart}
\AtBeginDocument{%
  \providecommand\BibTeX{{%
    \normalfont B\kern-0.5em{\scshape i\kern-0.25em b}\kern-0.8em\TeX}}}

\setcopyright{acmcopyright}
\copyrightyear{2018}
\acmYear{2018}
\acmDOI{XXXXXXX.XXXXXXX}

\acmJournal{JACM}
\acmVolume{37}
\acmNumber{4}
\acmArticle{111}
\acmMonth{8}

\usepackage[nolist,nohyperlinks, printonlyused, withpage]{acronym}
\usepackage{csquotes}

\begin{document}

\title{Designing for Disengagement: Challenges and Opportunities for Game Design to Support Children's Exit From Play}


\author{Meshaiel Alsheail}
\affiliation{%
  \institution{Karlsruhe Institute of Technology}  
  \city{Karlsruhe}
  \country{Germany}
}
\email{meshaiel.alsheail@kit.edu}

\author{Dmitry Alexandrovsky}
\affiliation{%
  \institution{Karlsruhe Institute of Technology}  
  \city{Karlsruhe}
  \country{Germany}
}
\email{dmitry.alexandrovsky@kit.edu}

\author{Kathrin Gerling}
\affiliation{%
  \institution{Karlsruhe Institute of Technology}  
  \city{Karlsruhe}
  \country{Germany}
}
\email{kathrin.gerling@kit.edu}

\renewcommand{\shortauthors}{Alsheail, et al.}

\begin{acronym}
\acro{HCI}{Human-Computer Interaction}
\acro{PX}{Player Experience}
\acro{SDT}{Self-Determination Theory}
\acro{UE}{User Engagement}
\acro{UX}{User Experience}
\end{acronym}

\begin{abstract}
Games research and industry have developed a solid understanding of how to design engaging, playful experiences that draws players in for hours and causes them to lose their sense of time. 
While these designs can provide enjoyable experiences, many individuals -- especially children -- may find it challenging to regulate their playing time, and often they struggle to turn off the game. 
In turn, this affords external regulation of children's playing behavior by limiting playing time or encouraging alternative activities, which frequently leads to conflicts between parents and the children.
Here, we see an opportunity for game design to address player disengagement through design, facilitating a timely and autonomous exit from play. 
Hence, while most research and practitioners design for maximizing player engagement, we argue for a perspective shift towards disengagement as a design tool that allows for unobtrusive and smooth exits from the game. We advocate that interweaveing disengagement into the game design could reduce friction within families, allowing children to finish game sessions more easily, facilitate a sense of autonomy, and support an overall healthier relationship with games.
In this position paper, we outline a research agenda that examines how game design can address player disengagement, what challenges exist in the specific context of games for children, and how such approaches can be reconciled with the experiential, artistic, and commercial goals of games. 
\end{abstract}



\received{20 February 2007}
\received[revised]{12 March 2009}
\received[accepted]{5 June 2009}

\maketitle

\section{Introduction}
Supporting disengagement from play is relevant for children as it contributes to healthy gaming practices, can help avoid conflict within families about playing time, and could potentially reduce the risk of harmful overuse of games. While there is a substantial body of literature that addresses the lack of disengagement from play through the lens of pathological use of games (e.g., clinical perspectives on game addiction~\cite{fisher1994, griffiths1997}, surprisingly little is known about disengagement from games from the perspective of games design, i.e., whether and how specific design strategies or game mechanics could support the exit from play.
Instead, much of the work addresses continued player engagement (both in the context of positive game design - for example, achieving flow~\cite{cowley2008}, but also with a critical view - for example, work on dark patterns~\cite{zagal2013}), and \ac{HCI} games research likewise overwhelmingly focuses on the entry into play (cf. ~\cite{petersen2017}) and continued participation in it(cf.~\cite{lewis2012, bowman2021, deleonpereira2021}). 
Where disengagement is addressed, it is often done so with a negative connotation through the lens of player attrition (e.g.,~\cite{alexandrovsky2021, hadiji2014, hui2013}), and from a restrictive perspective such as external terminating of play~\cite{stevens2021}. 
The negative lens of \ac{HCI} research on disengagement has previously been criticized by O'Brien and colleagues~\cite{obrien2022}, who outlined that disengagement can also be temporary, a result of momentary satisfaction with the preceding experience, and an expression of user agency. 
We argue that the limited view on disengagement is also  a missed opportunity for games research: Appreciating the final stages of play as a part of \ac{PX} that should actively be designed for gives game designers an additional tool within their box. 
Additionally, developing strategies that support players in the achievement of an exit from play at their own volition should contribute to player autonomy (which has for example been extensively studied in the context of remaining within play~\cite{deterding2016}), for younger and older players alike.

In this position paper, we reflect on the potential benefits of designing for disengagement in games for children, an audience that still establishes gaming habits and needs to negotiate these in a family context. To this end, we summarize current perspectives in \ac{HCI} games research on player engagement, we give an overview of current industry best practices designed to limit children's engagement with games, and we outline pathways to game design-driven strategies for player disengagement that empower children and their parents to exit play in a positive context.

\section{Disengagement, Digital Games, and Children's Engagement}
In this section, we first introduce how \ac{HCI} research defines disengagement. Then, we examine how the concept is approached in \ac{HCI} games research, and we reflect upon current best practices in industry and research to support children's disengagement from digital games and other media.

\subsection{Defining Disengagement within HCI Research}
Research in \ac{HCI} has approached engagement and disengagement from different perspectives. In this work, we follow \citet{obrien2008}'s definition of \acf{UE} as the theoretical foundation.
\ac{UE} refers to the \textquote[\cite{obrien2016}]{quality of user experience characterized by the depth of an actor’s investment when interacting with a digital system} and~\textquote[\cite{attfield2011}]{emphasises the positive aspects of the interaction and, in particular, the phenomena associated with being captivated by the technology}. 
As~\citet{attfield2011} stated, \textquote[\cite{attfield2011}]{successful technologies are not just used, they are engaged with; users invest time, attention, and emotion into the equation}.
Therefore, in most cases, it is desirable to have an engaging interactive system as it facilitates retention, productivity, and overall satisfaction.
The sense of engagement is inferred by cognitive, affective, and behavioral constructs such as flow~\citep{csikszentmihalyi1990}, motivation~\citep{rigby2011}, attention~\citep{attfield2011, obrien2008}, or adherence~\citep{couper2010, nacke2011, perski2017}.
The cognitive aspect of engagement frequently relies on conscious components such as attention, interest, or effort~\citep{doherty2019, islassedano2013,sun2014}.
The affective component of engagement encompass the subjective emotional responses including enjoyment, aesthetics, endurability, and novelty~\citep{obrien2008, doherty2019}.
Behavioral component describes the action and participation with the activity~\citep{doherty2019, islassedano2013, silpasuwanchai2016}.
\citet{obrien2008} conceptualize four stages of UE: \textit{point of engagement} is the first contact with the interactive system; \textit{period of sustained engagement} is the actual time span users interacting with the system; \textit{disengagement} describes the termination point of an engaging period (e.g., end of a session); and \textit{re-engagement} is referred to when users return to the interactive system and is marked by active choice taken from the user. Each of these phases is characterized by different attributes of the \ac{UX} that the interaction design should emphasize~\citep{obrien2008}. 
These four stages allow conceptualizing experiences with interactive systems on a timeline with interaction cycles where each stage gives specific target points for the interaction design that researchers and designers could use to orchestrate the experience (e.g., the behavior of the user).


\subsection{Engagement and Disengagement in Games Research}
\label{sec:engagement_disengagement_in_games}
In games research, engagement describes the act of players being fully absorbed and involved in the game, characterized by a high level of motivation, attention, and emotional investment \cite{schonau-fog2012, rapp2022}.
In the context of games, \ac{SDT} is commonly used to explain the motivation of play~\cite{deci2012}. \ac{SDT} suggests that providing players with opportunities to make choices that align with their interests and values, experience a sense of mastery or progress, and connect with other players or characters in the game will lead to increased engagement and enjoyment~\cite{tyack2020, tyack2020a}. In line with \ac{SDT}, Flow theory proposes that an optimal and thus, engaging experience emerges when the players' skill and game difficulty are well-matched~\cite{csikszentmihalyi2000}.
A significant and growing body of research has been conducted on what keeps players engaged with the game, for example, through game updates, by providing new content and boosting the experience for the better through challenges, which creates satisfaction and enhances the \ac{PX}~\cite{zhong2021,claypool2017,altimira2017,zhong2022}. 

\textbf{From a commercial perspective, publishers are interested in player engagement to increase profitability} through game subscriptions, attract and retain customers, and keep the game entertaining with new content~\cite{baptista2019,rooija.j.van2021}. In consequence, game designers increasingly use strategies with manipulative elements to keep users connected to their games~\cite{rooija.j.van2021,zagal2013}. Among others, such \textit{dark patterns} include temporal dark patterns (i.e., employ time-limited tactics to create a sense of urgency, encouraging users to take desired actions) and psychological manipulations (such as discounts on resources, encouraging users to invest money to enhance their gameplay) make it more challenging for players to disengage from play at their own volition \cite{zagal2013,rooija.j.van2021}.

Despite these concerning design strategies, \textbf{the dominant perspective in games research is that disengagement  -- the process in which the user retreats from interacting with a system either temporarily or permanently \cite{obrien2022} -- is the result of poor game design} that does not encourage continued player engagement \cite{obrien2022}. For example,~\citet{xie2014} studied how to predict player disengagement in online games in an effort to identify instances in which designers would need to address design flaws. 
Similarly, ~\citet{oertel2020, ben-youssef2021}  have presented disengagement in HCI as halting the problematic or meaningless consumption of technology after feelings of "frustration" based on poor design or lack of motivation in its design.



Generally, we need to acknowledge the \textbf{tension between the aforementioned \textit{dark patterns} and the player's ability to achieve a satisfying experience and retain agency to end play at a point in time that is convenient for them}. In consequence, there is a large body of literature focusing on problematic play behaviors (i.e., excessive play\cite{sublette2012}, and gaming addiction \cite{griffiths2009,wood2008, kuss2012}). Particularly in the context of children and their engagement with games, this is discussed through the lens of youth well-being~\cite{gil2020, vanrooij2017, khorsandi2022, elsayed2021} and frequently addressed through external strategies to support disengagement from play, which we discuss in the following section~\cite{clark2011,peters2018}.

\subsection{Current Approaches to Support Children’s Disengagement from Games}
\label{sec:children_disengagement}
Research and industry have explored a range of strategies to support children's disengagement from games and reduce playing time.

\textbf{The dominant strategy to address (the lack of) disengagement is through the introduction of time restrictions, either at an individual or societal level.} Such tools involve timers \cite{hiniker2016a}, trackers of usage \cite{kim2016a, RescueTime2007}, automated nudges to disengage \cite{okeke2018}, promoting self-regulation through social support and goal-setting \cite{ko2015}, or block users entirely from using the device or specific apps~\cite{lee2014, jasper2015}. Directly addressing children, various tools seek to manage screen time and other issues by allowing parents to set time limits \citet{bieke2016}, e.g., Net Nanny~\cite{ross2021}, CYBERsitter~\cite{milburn1998}, child-friendly filters on Netflix or Apple's ScreenTime\footnote{\url{https://support.apple.com/en-ca/HT208982}} which also provide detailed statistics about the usage of individual apps.

Time restrictions are also introduced at a societal level. For example, South Korea implemented in 2011 a law that regulates how much players are allowed to play within a 24-hour period. 
Likewise, China introduced a time-limit policy that decreases players' rewards after a play window of 3 hours~\cite{kiraly2018}. 

\textbf{Within games, such patterns have been discussed as \textit{blocking} and \textit{waiting} mechanics and are frequently reported as dark patterns} \cite{alexandrovsky2019, alexandrovsky2021}.
The approaches to reduce screen time in games can be explicit, such as the MS XBox warning players about excessive gameplay times~\cite{microsoft2016} or more subtle like in \textit{Stronghold: Crusader} every now and then, the in-game companion suggests the player take a break or asks if they want to drink~\cite{Crusader2002}. However, such disengagement strategies can reduce the players' sense of autonomy and can leave players with an unsatisfied experience feeding the wish to continue playing~\cite{davies2016}. Likewise, hindering players from achieving their goals can in fact cause frustration and aggression~\cite{card2011,battigalli2015}. 
However, planning out screen time in advance can help children to disengage from screen exposure~\cite{hashish2014,hiniker2018b}. Likewise,~\citet{zhang2022a} showed for social media usage that design patterns facilitating the users' agency are more effective than time-restricting methods.

For \textbf{children's media usage}, \citet{barr2020} noted that the context of media usage in families is rarely considered in the literature and argue that to understand the long-term effects of children interacting with digital media, research needs to consider measures beyond screen time. 
Here, parental mediation of media use is an important pillar, but parents often are challenged in assessing the risks the children might encounter with digital media~\cite{bieke2016,mitchell2005}. 
The Parental Mediation Theory categorizes three communication strategies -- active, restrictive, and co-viewing mediation -- that can be leveraged to mitigate the negative effects of media use.
Studies on parental mediation of violent TV consumption showed that both active and restrictive mediation has been negatively related to the children's aggressive tendencies, while co-viewing showed a positive influence on the child's aggression~\cite{nathanson1999}. 
This is in line with situated learning theories which suggest that children learn through \textquote[\cite{brown1989}]{cognitive apprenticeship} which transforms learning from a process of transition towards a meaningful social activity~\cite{clark2011}.
Hence, \citet{bieke2016} argue that parental protection of children's media usage should not result in "helicopter apps" but rather support discussions between parents and children and encourage the child's autonomy~\cite{clark2011}.

Here, some work exists on \textbf{children's consumption of video and TV that addresses disengagement beyond restrictions}. For example, \textit{Coco's Videos}~\cite{hiniker2018b} is a child-friendly video player that supports the child's self-regulation of screen exposure by letting the children decide how much time they like to spend watching. When the time has run out, a virtual character appears informing about the end of the session and making suggestions for other activities. The authors report the dialog with the virtual character at the end of the session gained value for the children and became part of the transitioning ritual. Likewise, \textit{FamiLync}~\cite{ko2015a} emphasizes both the parent's and children's sense of agency and provides support for a participatory and elucidating parental mediation of media consumption. Another approach promises to mitigate the side effects of screen-time restrictions using the physical periphery around the screen device to move the child's attention and ease the transition from immersion~\cite{yim2021}.

Reviewing the literature on disengagement, \textbf{we observe several gaps in research}: First, HCI and games research have only begun to address disengagement as part of the engagement cycle, mostly viewing it as a negative event. Second, common strategies to foster player disengagement view games as static objects, and instead seek to provide external strategies that help regulate the behavior of children and other players. Third, work that addresses children's media use has only begun to take into account developmental perspectives and family relationships, which leaves rooms for work in this space. 

\section{Challenges and Opportunities for Research: Supporting Disengagement While Maintaining Positive Play Experiences}
A key challenge for strategies to support disengagement from play is the tension between games seeking to provide immersive and engaging experience, while giving players -- including children -- autonomy to exit the experience at their own volition. Based on our examination of the perspective of games research on player disengagement, current best practices to facilitate disengagement of younger players, and the gaps therein, we would like to highlight the following three areas for future research.

\subsection{Developing an Evidence-Based Perspective on Children's Exit From Play}
Much to our surprise, little is known about how children experience the exit from play beyond literature addressing problematic gaming (cf. Section~\ref{sec:engagement_disengagement_in_games}), suggesting that children's perspectives on ending play are poorly understood. Likewise, industry best practices on limiting children's playing time (cf. Section~\ref{sec:children_disengagement}) are neither rooted in an understanding of children's cognitive development nor reflect what we know about engaging \acp{PX}. For example, a prominent strategy to support disengagement is to introduce a maximum playing time, after which the gaming experience automatically ends. However, we know that games as immersive artifacts affect player perception of time (e.g., when experiencing flow \cite{csikszentmihalyi2000}), and in the context of children, the issue is exacerbated by the fact that humans only develop a concept of time from the age of seven~\cite{droit-volet2013}. Here, we see potential in a two-prong research approach that first seeks to understand the specific experience that children have when exiting games, and then developing evidence-based strategies to support the exit from play that are rooted in an understanding of children's cognitive development. Thereby, we would assume that designers and researchers could achieve the implementation of strategies that reduce friction within families, while maintaining a more positive overall gaming experience.

\subsection{Accounting for the Child-Parent Relationship in the Disengagement Process}
The relationship between kids, parents, and games is complex, and should likewise be taken into account when designing the disengagement process. 
For example, \citet{donati2021} found that setting restrictions on the time, place, and content of video gaming can prevent excessive gaming, but \citet{papadakis2022} illustrates that parents struggle to control their children's time spent on tablets, with \cite{donati2021} suggesting that the effectiveness of such rules is moderated by the degree of parent-child agreement. 
\citet{kahila2022} report that children can experience intense anger when their in-game experiences are interrupted, for example, when parents remind them of homework, household chores, or meal times), especially when the game is going well. From the perspective of HCI games research, this presents an opportunity to design mechanics to support exit from play that account for the complexity of the relationship between parents and children, as well as family life. Instead of setting generic time limits, this could mean providing parents with the tools of understanding their children's experiences with games particularly addressing the question of when it is a 'good time' to quit, but also casting the process of shared responsibility in which parents need to support their children in finding an appropriate end, and researchers and designers need to weigh the needs of both groups.

\subsection{Appreciating Disengagement as Part of Play}
Currently, disengaging from games is underappreciated by HCI games research, and we believe that it is a research opportunity to view it as a natural part of play. Reflecting the four phases of the engagement cycle proposed by~\citet{obrien2008}, a point of engagement, a period of engagement, disengagement, and finally, re-engagement, we align with their conceptualization of disenagement as a natural part of the engagement--disengagement--re-engagement cycle, which is spanned across two dimensions: the degree of users' agency and the span between positive and negative engagement. This definition of disengagement encompasses the users' goals, the meaning of usage, and the degree of control the users take over interacting with the system.. Thereby, it becomes possible to design for specific exit experiences, and to consider concerns around disengagement and excessive play from a positive perspective, shifting the focus to player empowerment, enabling them to re-gain agency over the time at which they (temporarily) end their engagement with a specific game.
 Recently, \citet{stevens2021} examined the effectiveness of currently implemented design strategies in the context of overuse of digital games, and concluded that features that set limits on playtime or locked players out of the game received low support (65\% disapproval) among habitual and problem gamers \cite{stevens2021}. While not primarily examining disengagement, a study by ~\citet{tyack2020a} found that games that allow players to pause or save their progress and come back later, without losing their progress or rewards, can help players to disengage without feeling frustrated or stressed~\cite{tyack2020a}. Also, ~\citet{alharthi2018} point out that idle games, which reward players for waiting, can be an effective way to foster disengagement without harming the \ac{PX}~\cite{alharthi2018}. Similarly, \citet{davies2016} report that players who regarded their play session as completed (i.e., achieving their goals) when blocked from playing felt less frustrated than those who were cut out in the middle of a quest.
 This highlights the need for features that extend beyond extrinsic control, examining how to make it easier for players to quit games at their own volition, and to create games that have natural end points. This is in line with 
~\citet{obrien2008}, who theorizes that disengagement could be related to positive emotions such as feeling successful and satisfied when achieving a goal, which is something that we hope games research and game design can aspire to.

\section{Conclusion}
Within the HCI games research community, disengagement is an underresearched aspect of player \textbf{en}gagement with games, and is either framed as a result of poor game design, or examined in the context of problematic overuse of games. The latter also is a prominent perspective on children's disengagement from games, which is predominantly addressed through external means of regulating playing time (e.g., built-in time limits and parental mediation). In our position paper, we argue that this is a missed opportunity for game design to expand beyond restrictive practices. Here, a central question that remains for our research community is whether (the lack of) disengagement should be addressed through external regulation, or whether we can shift toward a perspective where games researchers and designers actively design for player disengagement, enabling players of all ages to establish healthy relationships with their favourite medium.

\begin{acks}
\end{acks}

\bibliographystyle{ACM-Reference-Format}
\bibliography{references}

\end{document}